\documentclass[10pt,english]{article}
\usepackage[T1]{fontenc}
\usepackage[latin2]{inputenc}
\usepackage{geometry}
\usepackage{babel}
\usepackage{graphics}
\usepackage{floatflt}

\makeatletter

\providecommand{\LyX}{L\kern-.1667em\lower.25em\hbox{Y}\kern-.125emX\@}
\let\SF@@footnote\footnote
\def\footnote{\ifx\protect\@typeset@protect
    \expandafter\SF@@footnote
  \else
    \expandafter\SF@gobble@opt
  \fi
}
\expandafter\def\csname SF@gobble@opt \endcsname{\@ifnextchar[
  \SF@gobble@twobracket
  \@gobble
}
\edef\SF@gobble@opt{\noexpand\protect
  \expandafter\noexpand\csname SF@gobble@opt \endcsname}
\def\SF@gobble@twobracket[#1]#2{}

 \newcommand{\lyxaddress}[1]{
   \par {\raggedright #1 
   \vspace{1.4em}
   \noindent\par}
 }

\makeatother

\begin{document}

\title{Automatic variability analysis of bulge stars in OGLE II image subtraction
database}

\author{\textbf{Tomasz Mizerski and Michał Bejger}}

\maketitle

\lyxaddress{\centering Warsaw University Observatory, Al. Ujazdowskie 4, 00-478 Warszawa,
Poland}

{\par\centering mizerski, bejger@astrouw.edu.pl\par}

\begin{abstract}
We present results of star variability analysis in OGLE II first bulge field.
Photometric database was derived by means of image subtraction method (Woźniak
2000) and contains 4597 objects pre-classified as variables. We analyzed all
the light curves in order to find periodic variables, non periodic but all-time
variables and stars showing episodic changes in their brightness, e.g. gravitational
lenses. Variability of 3969 stars was confirmed, among them we detected 12 lensing
event candidates whose light curves are shown. We found 762 periodic variables.
Algorithmic methods let us identify 71 RR Lyrae and 110 W UMa stars. Almost
all luminous red giants are found to be variable. Most of red clump giants are
not variable. Classification of all 4597 objects is presented online. 
\end{abstract}

\section{Introduction }

The observational data were obtained with Warsaw 1.3m telescope located at Las
Campanas Observatory of the Carnegie Institution of Washington, and instruments
(\char`\"{}Zero generation\char`\"{} single chip, camera SITe 2048\( \times  \)2048
thin chip) dedicated to the second phase of the Optical Gravitational Lensing
Experiment - OGLE II. In the years 1997 - 1999 about 200 observations were made,
mainly in drift scan mode in the I photometric band. Single image is 8192x2048
pixels (55'x14') in size - we refer to Udalski, Kubiak \& Szymański (1997) for
the technical details of instruments and the telescope. 

The photometric data based on the algorithm of Alard \& Lupton (1998) and Alard
(2000) and derived by P.R.Woźniak (Woźniak 2000) can be downloaded from \emph{http://astro.princeton.edu/\~{}wozniak/dia.}

The article is arranged in the following order: Section 2 describes our method
of rejecting spurious variables. In section 3 we describe the method and results
of periodic variability analysis. Afterwards, in section 4, search for specific
variability types - RR Lyrae and W UMa stars - is reviewed. Section 5 is dedicated
to non periodic variables. We discuss methods of detecting such stars - as an
example, simply statistical algorithms let us identify 12 gravitational lensing
events, in one case with lensed star being irregular variable itself. At the
end of the article, in section 6, we present color-magnitude diagram for detected
variables and demonstrate the luminosity function of the field.

\section{Spurious variables}

\begin{figure}[h]
{\par\centering \resizebox*{16cm}{16cm}{\includegraphics{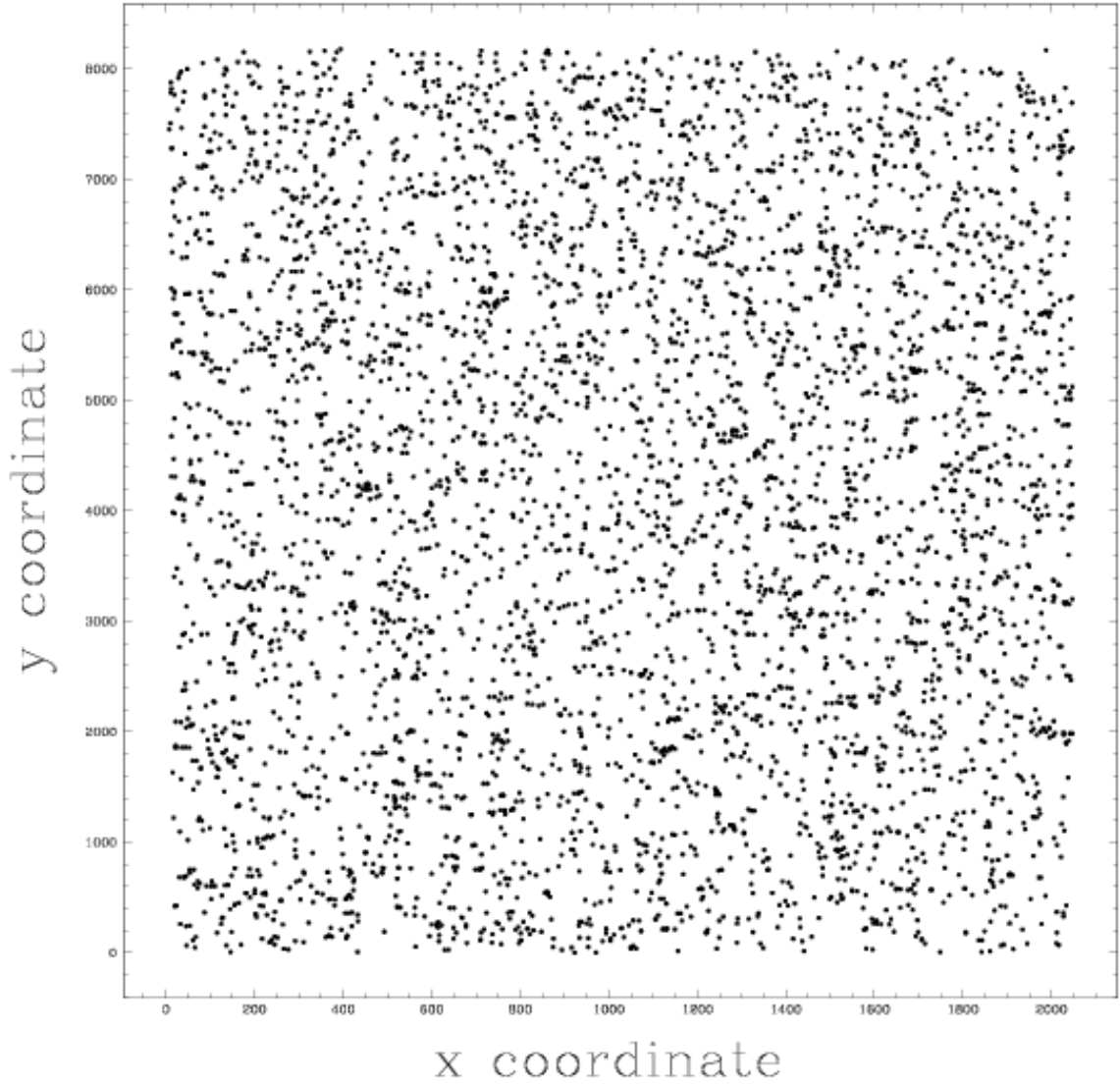}} \par}

\caption{{\small \label{coor}Coordinate plane of the reference image. Note the groups
of stars - in many cases a bright star influences its faint neighbors and produces
spurious variable stars.}\small }
\end{figure}

\begin{figure}[h]
{\par\centering \resizebox*{0.8\textwidth}{0.95\textheight}{\includegraphics{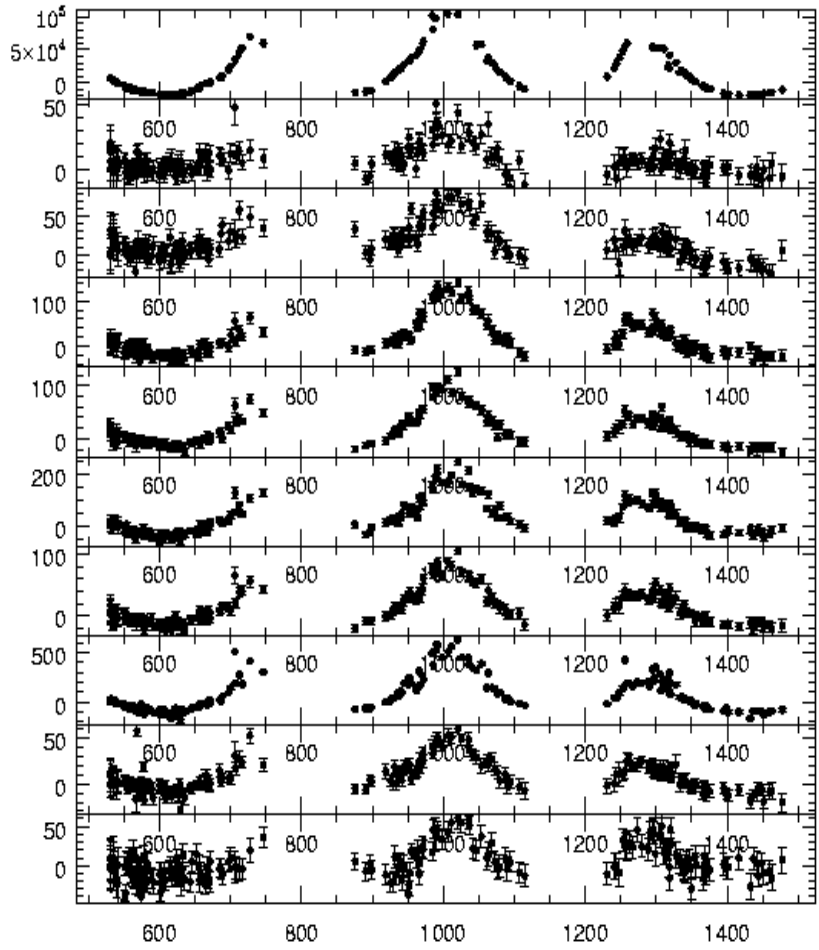}} \par}

\caption{{\small \label{spur}Real variable (top window) and nine fake variables located
in its vicinity.}\small }
\end{figure}

\begin{figure}[h]
{\par\centering \resizebox*{10cm}{10cm}{\includegraphics{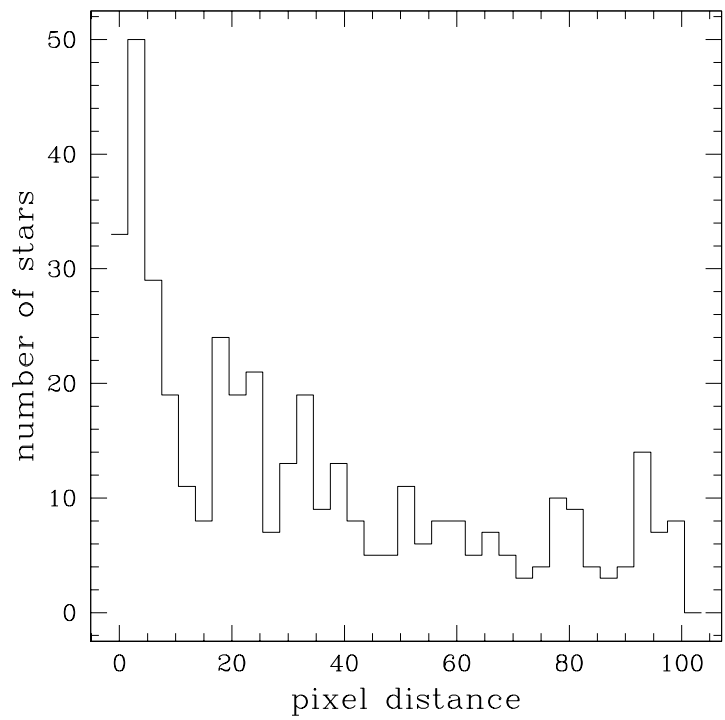}} \par}

\caption{{\small \label{spurhist}Histogram of distances between correlated genuine
bright variables and spurious variables.}\small }
\end{figure}

Some objects pre-classified as variables may not in fact be truly variable.
Apparent and especially temporary changes in brightness can be caused by instrumental
defects or by the method itself. If such stars tend to gather in groups in the
imaged field their variability is very improbable. In Fig. \ref{coor} we demonstrate
locations of all stars on the reference image coordinate plane. As we can see
correlations between positions of some objects are noticeable. By investigating
all dense regions we found groups of artificial variables. In the case illustrated
on Fig. \ref{spur} one very bright variable influenced nine of its faint constant
neighbors and produced nine spurious variable stars. To detect all such cases
we performed following procedure for every variable star: within the 100 pixel
radius we searched for stars whose light curves were correlated with the light
curve of the star. If the correlation coefficient was greater than 0.75 the
stars were considered correlated and thus only the brightest one could be regarded
as a real variable. In Fig. \ref{spurhist} we show histogram of distances between
luminous variable stars and spurious variables produced by them. Choosing the
radius of 100 pixels may result in finding few stars correlated by chance but
ensures that most of the spurious variables are rejected.

As a result we found 254 groups of correlated light curves and 363 stars were
classified as spurious variables. This problem concerns objects of all possible
variability types. Setting the correlation coefficient limit lower than 0.75
excludes real variables. 

Independently, stars exhibiting high proper motions were also taken into account.
Those stars produce pairs of anti-correlated light curves. There are 98 such
pairs. For more details on stars with high proper motions we refer to Eyer \&
Woźniak (2001). 

This improved classification decreased the number of variable candidates from
4597 to 3969 objects.

\section{Periodic variables. Criteria of acceptance}

The OGLE II database of the first bulge field bul\_sc1 (\( \alpha _{2000}=18\mathrm{h}\, 02\mathrm{m}\, 32.5\mathrm{s},\: \beta _{2000}=-29^{\mathrm{o}}57'41'',\: l=1.08^{\mathrm{o}},\: b=-3,62^{\mathrm{o}}) \)
consists of about 200 frames from three observational seasons. The time span
\( \Delta t \) is \( \sim  \) 1000 days. Two periodic signals with frequencies
that differ by \( \Delta \nu  \) will after \( \Delta t \) end up having phase
difference \( \Delta \varphi  \) = \( 2\pi \Delta \nu \cdot \Delta t \). We
demand \( \Delta \varphi  \) \( \sim  \) \( 10^{-2} \) which gives \( \Delta \nu  \)
\( \sim  \) \( 10^{-5}. \) With such a frequency step periodogram generation
takes a lot of CPU time if we search for short periods, i.e. high frequencies.
The only way to avoid this is to have a shorter time span \( \Delta t \). We
decided to search for short periodic variables using only the data points from
the first season. There are about 100 such points and \( \Delta t \) is only
200 days. This allows setting a greater frequency step in periodogram generator,
namely \( 10^{-4} \) {[}\( \mathrm{d}^{-1} \){]}. Time span of about 200 days
is robust enough for detecting variables with periods shorter than 50 days -
the most CPU consuming part. Variables with periods longer than 50 days were
searched for on the whole data set with frequency step equal to \( 10^{-5} \)
{[}\( \mathrm{d}^{-1} \){]}. 

We used Analysis of Variance algorithm (Schwarzenberg-Czerny 1989) to obtain
probable periods for every star. If such periods were found next step was to
check if any of them is physical. Using the usual \( \chi ^{2} \) minimization,
for each period a Fourier series was fitted to the data points. Number of harmonics
was set to six which is good enough to approximate most of the light curve shapes.
The purpose was not to refine the period but to find the coefficients of the
fit, that is amplitudes of all harmonic terms. The formula for \( \chi ^{2} \)
is given below: 

{\par\centering 
\begin{equation}
\chi ^{2}=\frac{1}{n-b}\cdot \sum ^{n}_{i=1}\frac{1}{\sigma ^{2}_{i}}\cdot \left( f_{i}-\sum ^{m}_{j=0}(A_{j}\cdot \sin (j\omega t_{i})+B_{j}\cdot \cos (j\omega t_{i}))\right) ^{2}
\end{equation}
\par}

where \emph{\( n \)} is the number of data points, \( m \) is equal to the
number of harmonics, \emph{\( b \)} is equal to the number of degrees of freedom
used for fit calculation; \emph{\( b=2m+1 \)}, \( f_{i} \) is the value of
flux on the i-th frame, \( t_{i} \) represents time of the i-th observation,
\( \sigma _{i} \) is the corresponding flux error and \emph{\( A_{j} \)} and
\( B_{j} \) are the harmonic terms amplitudes. Once an analytical approximation
of the curve was found it was possible to determine the quality of the fit.
Our parameter is defined as follows:

{\par\centering 
\begin{equation}
\rho =\frac{\frac{1}{n-b}\cdot \sum ^{n}_{i=1}\left( f_{i}-\sum ^{m}_{j=0}(A_{j}\cdot \sin (j\omega t_{i})+B_{j}\cdot \cos (j\omega t_{i}))\right) ^{2}}{\frac{1}{n-1}\cdot \sum _{i=1}^{n}\left( f_{i}-\frac{1}{n}\cdot \sum ^{n}_{j=1}f_{j}\right) ^{2}}
\end{equation}
\par}

\vspace{0.3cm}
Parameter \( \rho  \) measures dispersion from the fit and, contrary to \( \chi ^{2} \),
does not prefer low amplitude objects. For two sinusoidal curves with the same
period but different amplitudes, \( \chi ^{2} \) is significantly different,
while \( \rho  \) stays very much the same. For true periods it is mostly on
the order of \( 10^{-3} \) - \( 10^{-1} \) while spurious periods give relatively
much higher values of \( \rho  \).

Next thing to do was to find \( \rho  \)\( _{\mathrm{max}} \) - the upper
limit for the value of \emph{\( \rho  \)} as the criterion of acceptance. For
every star a list with possible periods and corresponding \( \rho  \)'s was
produced. We then sought the smallest \( \rho  \) on the list. The star was
considered periodic variable, when the value of \( \rho  \) was smaller than
the limit of acceptance i.e. the value of \( \rho _{\mathrm{max}} \). One should
not expect one universal value for \emph{}\( \rho _{\mathrm{max}} \): note,
that it is easier to get a low \( \rho  \) by chance when number of data points
\emph{}is smaller. That's why \( \rho _{\mathrm{max}} \) should rather be a
monotonically increasing function of \emph{\( n \)}. For the first season values
of \emph{\( n \)} are between 40 and 90. We chose 22 stars with different number
of data points in this range. 

\begin{figure}[h]
{\par\centering \resizebox*{10cm}{10cm}{\includegraphics{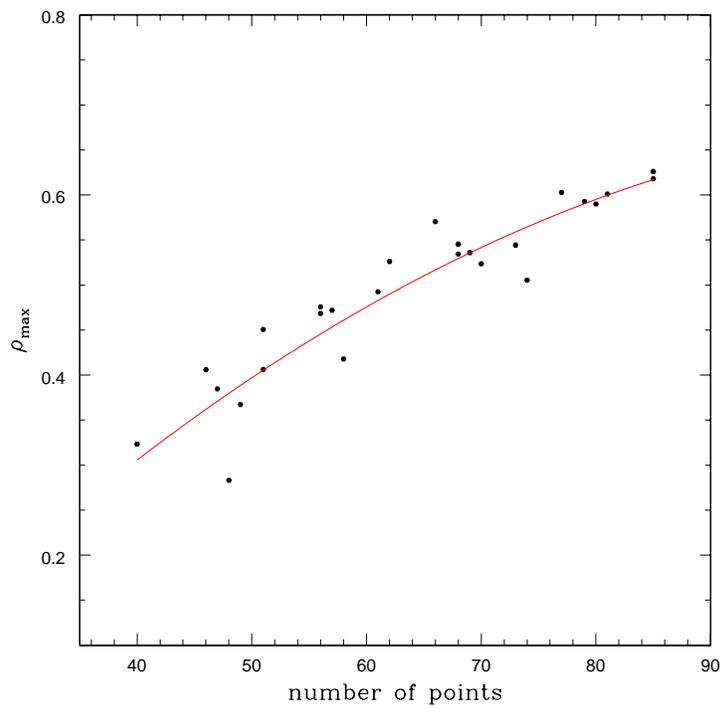}} \par}

\caption{{\small \label{rho}Dependence of \protect\( \rho _{\mathrm{max}}\protect \)
on the number of stars n. \protect\( \rho <\rho _{\mathrm{max}}\protect \)
is required for a star to be considered a periodic variable.}\small }
\end{figure}

For each of those 22 stars we generated 100 different artificial light curves
by shuffling their data points with corresponding errors. In the next step we
obtained suspected periods for all the generated curves and for each we chose
the period with the smallest \( \rho  \). As the result we got 22 lists with
100 smallest \emph{\( \rho  \)} each. For every list, which as we remember
corresponds to a different \emph{n}, we chose \( \rho _{\mathrm{max}} \) so
that only 2 out of 100 artificial stars would a have smaller \( \rho  \). We
arbitrarily defined \( \rho _{\mathrm{max}} \) as the mean value of the second
and the third smallest \( \rho  \). Finally we approximated the relation between
\emph{}\( \rho _{\mathrm{max}} \) and \emph{\( n \)} with parabola using least
squares. The result is shown in Fig. \ref{rho}. Having an analytical function
\( \rho _{\mathrm{max}}(n) \), we were able to determine \( \rho _{\mathrm{max}} \)
for any \emph{}\( n \). Our prescription implies that we allow about 2 per
cent spurious stars into periodic variable class. The goal is not to miss very
noisy periodic stars. 

The last step of verification was the same \emph{\( \rho  \)} test but on the
other two remaining observational seasons. The stars that had passed all these
tests were considered periodic variables. This procedure in the case of multi
periodic behavior would produce only one period - the better fitted one. For
example first overtone for RR Lyrae type d stars would be considered as the
physical period. It is not of a great importance since the star would still
be classified as a periodic variable.

\subsection{Results}

Using these techniques we found 487 periodic variables with periods shorter
than 50 days. The last step of verification - \( \rho  \) test on the other
remaining observational seasons turned out to be very important. Periodic variability
of only 487 of 1157 stars considered periodic variable on the first season was
confirmed. There are 265 stars with \( P\leq 1[d] \), 112 stars with \( 1[d]<P\leq 10[d] \)
and 110 stars with \( 10[d]<P\leq 50[d] \) among periodic variables detected
in this field.

Those are pulsating, eclipsing and miscellaneous type stars. Miscellaneous class
contains chromospherically active giants, sub giants and ellipsoidal variables
as well as all uncertain cases. Most of the eclipsing binaries are Algol and
W UMa type variables. Pulsators are mainly RR Lyrae and High Amplitude \( \delta  \)
Scuti (HADS) stars. Some examples are presented in Fig. \ref{per}. 

\begin{figure}[h]
{\par\centering \resizebox*{15cm}{15cm}{\includegraphics{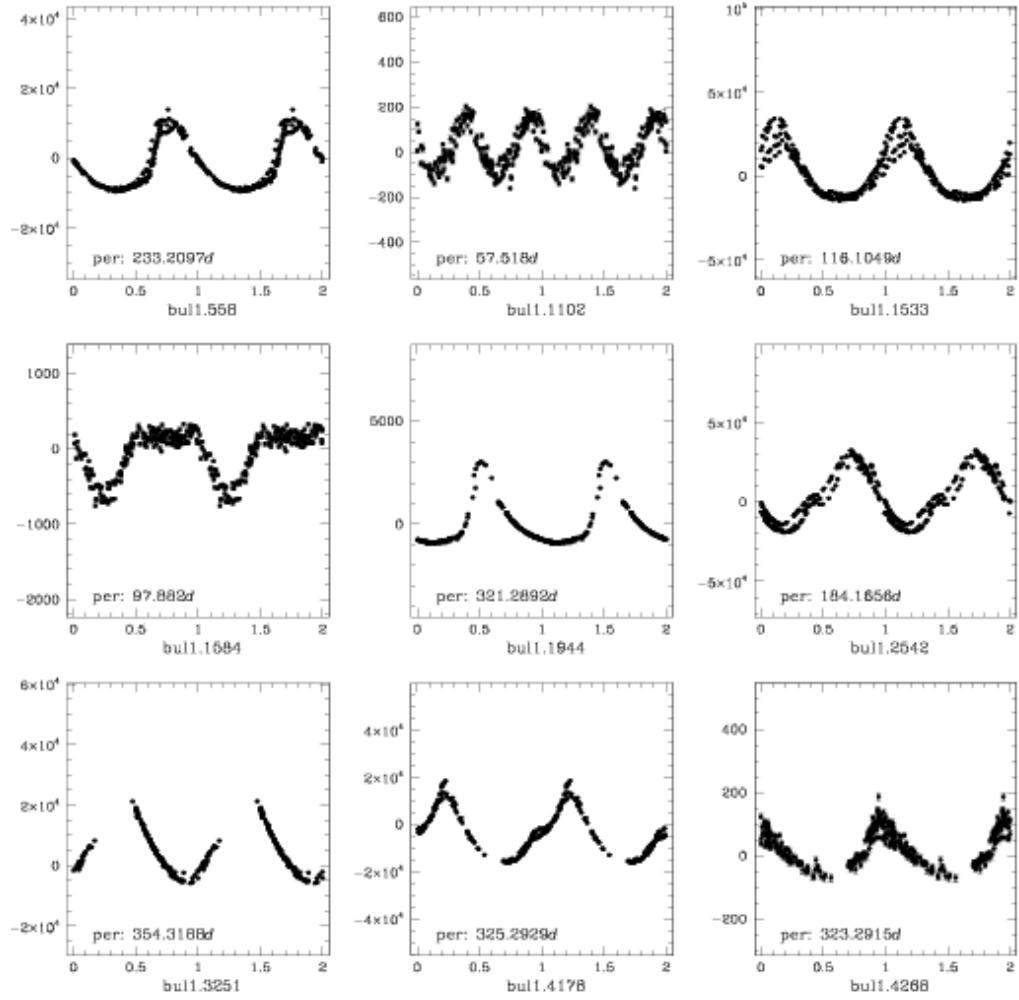}} \par}

\caption{{\small \label{long}A sample of long periodic variables. Difference flux is
plotted against phase.}\small }
\end{figure}

We also sought variables with periods longer than 50 days - 275 objects exhibit
strictly or more or less quasi periodic behavior. Examples are shown in Fig.
\ref{long}.

\section{Algorithmic methods for identification of common variability types}

\begin{figure}[h]
{\par\centering \resizebox*{15cm}{14cm}{\includegraphics{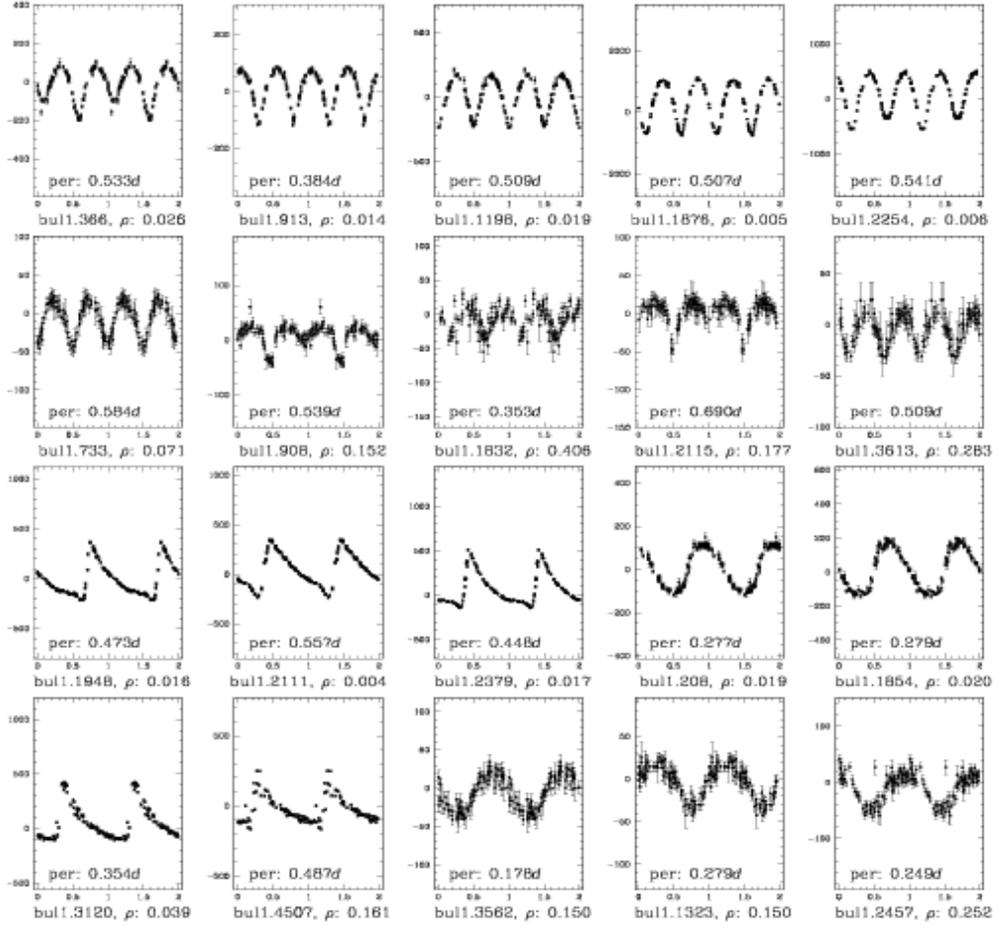}} \par}

\caption{{\small \label{common}We present a few light curves of W UMa and RR Lyrae
stars. There are well fitted ones, as well as lower signal to noise curves.}\small }
\end{figure}

After we had found periodic variables, we sought specific types of variability:
RR Lyrae and W UMa. There are well known methods of identifying these types
by means of Fourier series coefficients. For RR Lyrae we refer to Alard (1996)
and for W UMa to Ruciński (1993). Our code, based on those methods, but with
slightly modified parameters, detected 47 RRab, 24 RRc and 110 W UMa stars.
Apart from RR Lyrae our algorithm returned also 11 possible HADS stars. That
is because separating HADS and RRc is not always simple. Their light curve shapes
are sometimes similar to those of RR Lyrae type c stars. The main goal of the
method was to detect the types of variability by ``looking'' at the curve
shape only. In case of HADS stars we needed some additional parameters (like
luminosity) to make a choice. In Fig. \ref{common} we present sample light
curves with different dispersions. In the last row three last stars are HADS
stars candidates. The judgment is very problematic: their amplitudes are much
smaller than those of typical RR Lyrae type c, their Fourier coefficients however
are more RRc alike and the periods are longer than 0.2 d. Thus, we would like
to leave the ultimate judgment to the reader.

\section{Non periodic variable stars}

\subsection{Variance analysis}

\begin{figure}[h]
{\par\centering \resizebox*{0.75\textwidth}{0.57\textheight}{\includegraphics{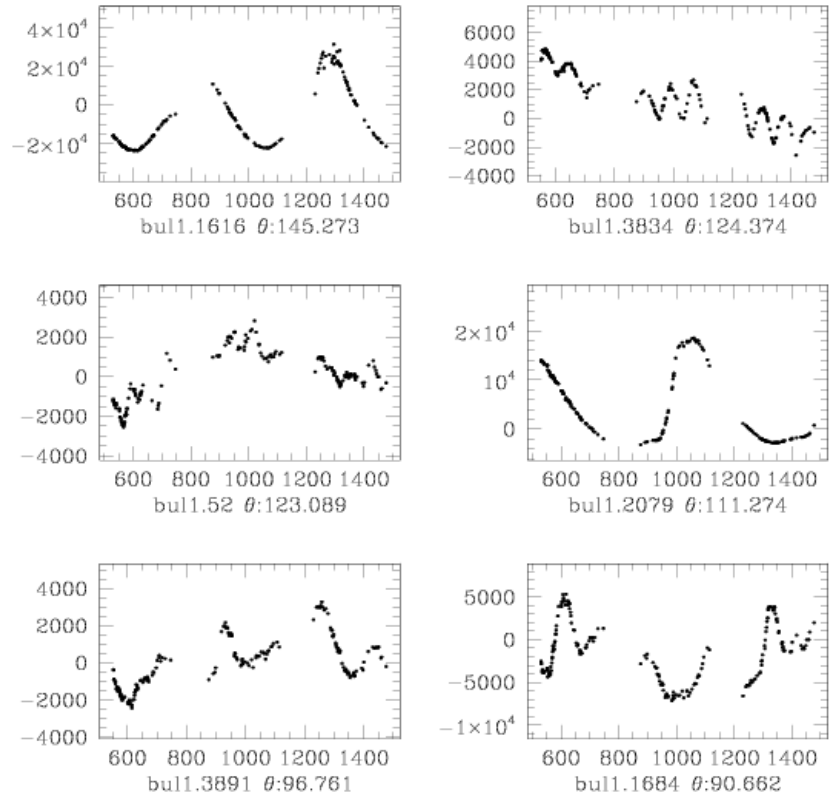}} \par}

\caption{{\small \label{aov}Examples of variables detected using Analysis of Variance.
Difference flux is plotted against time (in days)}\small }
\end{figure}

We used Analysis of Variance (AoV) in very much the same manner as Brandt (1970)
described. Whole data set was divided into packets, 10 observations each, and
mean value, as well as variance, was computed in every packet. Then we calculated
the variance of those mean values and mean packet variance. Their ratio \( \theta  \)
is an officious variability indicator since it tests mean values equality hypothesis.
As one can see in Fig. \ref{aov}, this algorithm works well for irregular behavior
detection. A constant star has \( \theta \sim 1 \) while for most of the microlensing
events found in this field \( \theta  \)\( \in [10,\, 100] \). 1341 stars
had their \( \theta >10 \), and this value seems to be a reasonable cut-off.
Weak lenses failed this test, but were detected by other means - as described
in section 5.3.

\subsection{Hunting noisy variables}

\begin{figure}[h]
{\par\centering \resizebox*{10cm}{10cm}{\includegraphics{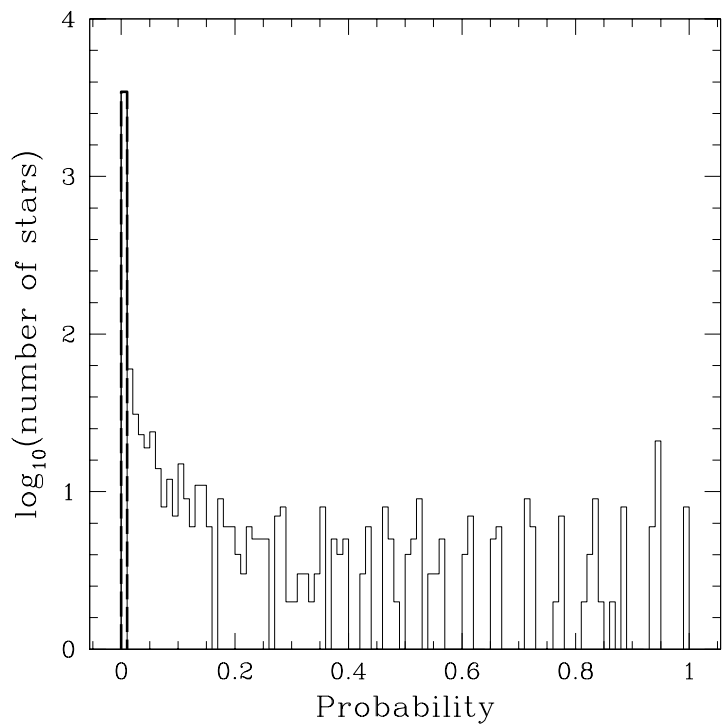}} \par}

\caption{{\small \label{hunt}Number of stars is plotted against the probability \protect\( P^{K}_{N}\protect \).
Dashed line indicates stars considered as variables by this test.}\small }
\end{figure}

Some stars exhibit a noisy behavior. To distinguish between a weak variable
star and photon noise, and to find all the cases where subsequent data points
are correlated, we developed the following test. Each data point was regarded
as positive/negative if it was greater/smaller than the weighted average of
the whole curve. Such treatment produces a time series of points, marked as
1 or -1 respectively. Along with time there are positive-negative state transitions,
i.e. sign changes, which have a binomial distribution. In the case of photon
noise this distribution is symmetric, since any point has equal probabilities
of being smaller or greater than the average \( N/2 \), where \( N \) is the
number of possible sign changes. For each light curve we compared its own distribution
of sign changes with the binomial distribution i.e. we calculated the probability
\begin{equation}
P^{K}_{N}(\left| N/2-k\right| \geq \left| N/2-K\right| )
\end{equation}
 where \( k \) is the number of sign changes in \( N \) possibilities, and
\( K \) equals the observed number of sign changes. We classified as variables
all the light curves with \( P^{K}_{N}\leq P(\left| N/2-k\right| \geq 3\sigma ) \),
with binomial distribution standard deviation \( \sigma  \)\( =\sqrt{N}/2 \).
There are 3447 such objects in the database, which leads to conclusion, that
522 stars were classified as non-variables by this test. Fig. \ref{hunt} shows
the histogram of computed probabilities against the logarithm of the number
of stars. Most of the rejected stars are stars with short time scales of variability.
One day sampling of the OGLE data can randomize flux distribution as far as
this test is considered. It does not affect long time scale variables since
their light curves are sampled with frequencies sufficient to reveal night to
night correlations. The short time scale variables, such as RR Lyrae or W UMa
stars were classified as variables by other means.

\subsection{Search for episodic variability}

Episodic variables such as gravitational lenses or long period eclipsing binaries
are easy to found using very simple statistical methods (of course, periodic
behavior was mainly detected by the use of methods described in section 3).
We rejected 10\% points with greatest absolute values of flux to calculate the
average value and dispersion. After those quantities were found we analyzed
each point`s deviation from the average value and counted separately all those
cases where the absolute value of deviation was 

{\par\raggedright 1) greater than \( 3\sigma  \) but smaller than \( 6\sigma  \)
(number of these points: \( n_{3} \))\par}

{\par\raggedright 2) greater than \( 6\sigma  \) but smaller than \( 9\sigma  \)
(\( n_{6} \))\par}

{\par\raggedright 3) greater than \( 9\sigma  \) (\( n_{9} \))\par}

{\par\raggedright The star was considered episodic variable if it satisfied at
least one of these arbitrary chosen criteria: \par}

{\par\raggedright 1) n\( _{3} \) \( \geq  \) 6\par}

{\par\raggedright 2) n\( _{6} \) \( \geq  \) 3 \par}

{\par\raggedright 3) n\( _{9} \) \( \geq  \) 3\par}

{\par\raggedright 4) n\( _{9} \) \( \geq  \) 2 and n\( _{6} \) \( \geq  \)
2\par}

{\par\raggedright 5) n\( _{9} \) \( \geq  \) 1 and n\( _{3} \) \( \geq  \)
3\par}

{\par\raggedright 6) n\( _{6} \) \( \geq  \) 1 and n\( _{3} \) \( \geq  \)
3\par}

We found 1710 such stars, some of them being not really episodic, but undoubtedly
variable, e.g. Mira-type stars. We also calculated the value of a parameter
defined as follows:

{\par\centering 
\begin{equation}
p=\frac{1}{3\cdot n_{3}+6\cdot n_{6}+9\cdot n_{9}}
\end{equation}
\par}

\begin{figure}[h]
{\par\centering \resizebox*{0.8\textwidth}{0.6\textheight}{\includegraphics{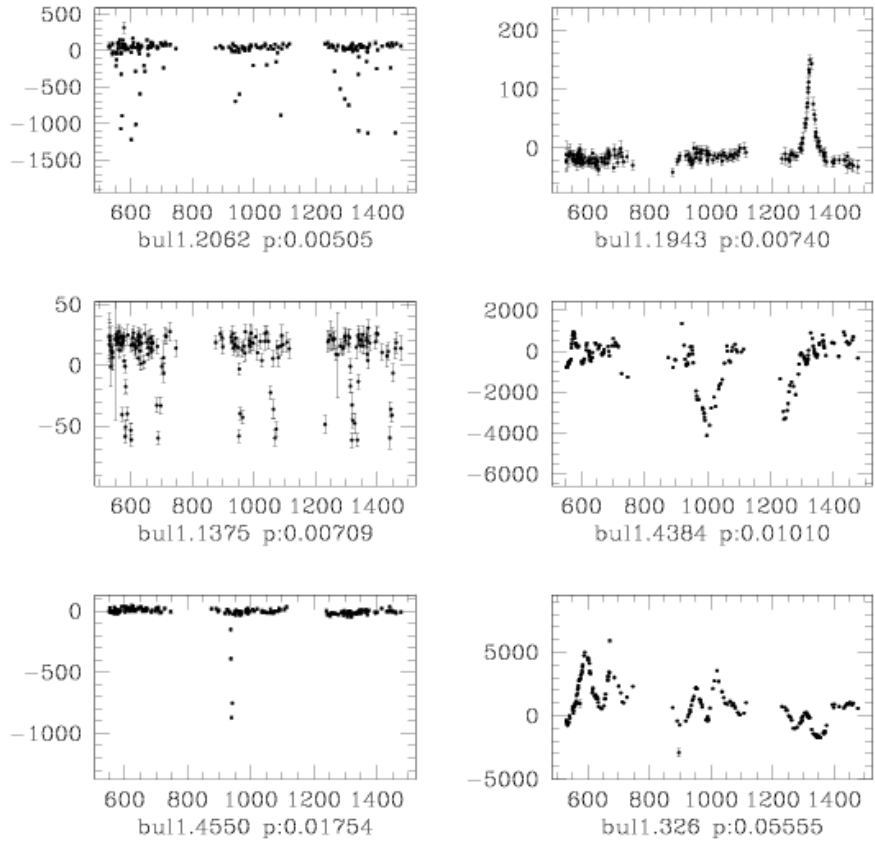}} \par}

\caption{{\small \label{episod}Sample stars found in episodic variable search. Values
of parameter p defined in this section are given.} \emph{}}
\end{figure}

This parameter is selective: it prefers points with greater deviations. Among
the stars with smallest values of p is a gravitational lens \emph{}bul\_sc1.1943
\emph{}which is shown on Fig. \ref{episod} along with other examples. This
strong lensing event had not been detected previously with the use of DoPhot.
In section 6 we give a possible explanation for this.

\subsection{Results. Gravitational lenses}

\begin{figure}[h]
{\par\centering \resizebox*{9cm}{9cm}{\includegraphics{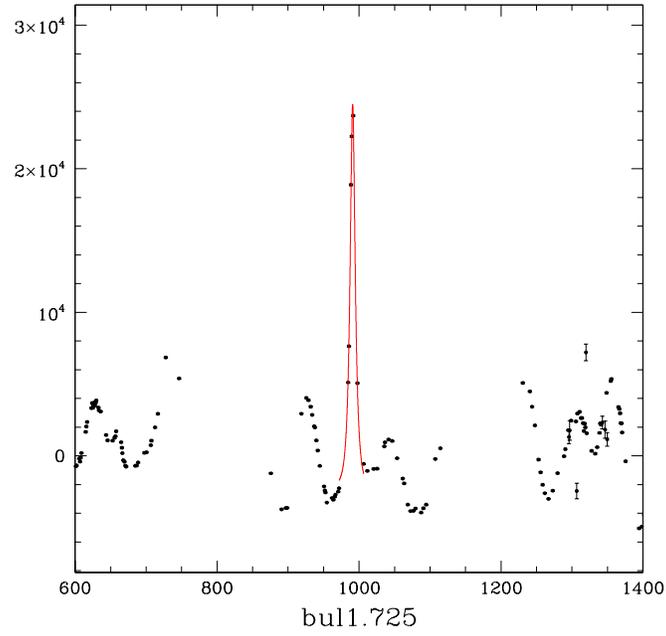}} \par}

{\par\centering \resizebox*{9cm}{9cm}{\includegraphics{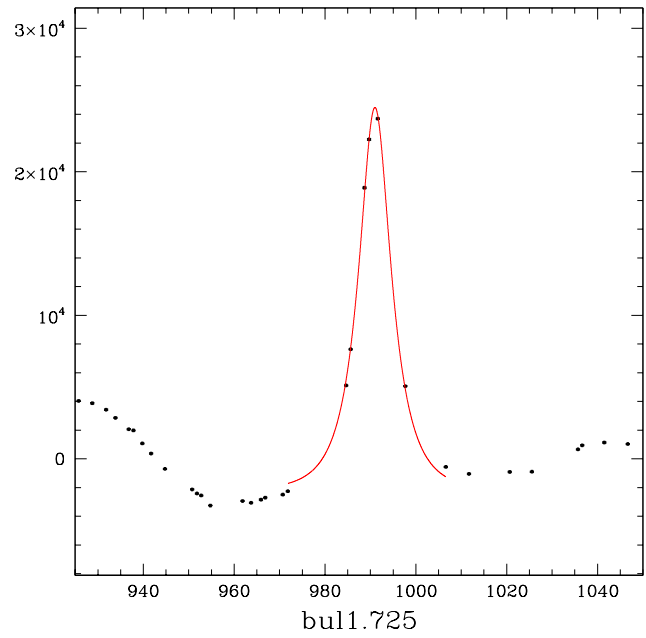}} \par}

\caption{{\small \label{725}The whole curve and close-up of} \emph{\small }{\small variable
gravitational lens candidate bul1.725. }\small }
\end{figure}

About 3950 stars passed at least one of the complementary tests described above.
In this number we identified 12 gravitational lens candidates. Woźniak et al.
(2001) reports 2 lenses in this field, that is bul\_sc1.1943 and bul\_sc1.2186.
All the others with exception for bul\_sc1.3061 and bul\_sc1.725 are in their
``transient'' catalogue. This results from stronger criteria used by those
authors. In table 1 we present theoretical models fitted to the data. Purely
mathematical approach to the fitting procedure may result in unphysical values
of some parameters because of degeneracy of the model curve. In the case of
\emph{}bul\_sc1.725, presented in Fig. \ref{725}, lensed star is variable itself,
thus there is no point in calculating a theoretical fit to the whole curve.
However, we show a fit to the data around the point of maximum magnification. 

\vspace{0.3cm}
{\centering \begin{tabular}{|l|l|l|l|l|l|l|l|l|l|l|l|}
\hline 
\textbf{\footnotesize star}{\footnotesize }&
{\footnotesize 1943}&
{\footnotesize 2186}&
{\footnotesize 1827}&
{\footnotesize 2096}&
{\footnotesize 2948}&
{\footnotesize 3061}&
{\footnotesize 2411}&
{\footnotesize 2724}&
{\footnotesize 3844}&
{\footnotesize 3087}&
{\footnotesize 3327}\\
\hline 
\hline 
{\footnotesize \( \chi ^{2} \)}&
{\footnotesize 1.03}&
{\footnotesize 1.27}&
{\footnotesize 1.03}&
{\footnotesize 1.49}&
{\footnotesize 1.04}&
{\footnotesize 1.16}&
{\footnotesize 1.30}&
{\footnotesize 1.51}&
{\footnotesize 1.33}&
{\footnotesize 1.27}&
{\footnotesize 1.07}\\
\hline 
{\footnotesize \( \mathrm{t}_{\mathrm{max}} \) {[}d{]}}&
{\footnotesize 1323.48}&
{\footnotesize 989.86}&
{\footnotesize 1454.61}&
{\footnotesize 1080.95}&
{\footnotesize 901.76}&
{\footnotesize 894.31}&
{\footnotesize 602.83}&
{\footnotesize 989.97}&
{\footnotesize 625.92}&
{\footnotesize 967.72}&
{\footnotesize 601.70}\\
\hline 
{\footnotesize \( \mathrm{t}_{0} \) {[}d{]}}&
{\footnotesize 26.13}&
{\footnotesize 25.88}&
{\footnotesize 39.15}&
{\footnotesize 19.14}&
{\footnotesize 31.13}&
{\footnotesize 15.27}&
{\footnotesize 10.92}&
{\footnotesize 1071.71}&
{\footnotesize 8.32}&
{\footnotesize 2.51}&
{\footnotesize 148.64}\\
\hline 
{\footnotesize \( u_{0} \)}&
{\footnotesize 1.29}&
{\footnotesize 0.23}&
{\footnotesize 1.29}&
{\footnotesize 0.51}&
{\footnotesize 0.46}&
{\footnotesize 0.56}&
{\footnotesize 0.15}&
{\footnotesize 0.0014}&
{\footnotesize 0.50}&
{\footnotesize 1.69}&
{\footnotesize 0.11}\\
\hline 
\end{tabular}\footnotesize \par}
\vspace{0.3cm}

{\par\centering Table 1: \textbf{}\( \chi ^{2} \) values and computed parameters
for eleven gravitational lensing event candidates\par}

The remaining 11 candidates also seem realistic since in each case scatter around
the theoretical fit, described by \( \chi ^{2} \) per degree of freedom, is
\( \sim  \) 1 (see table 1 for details). However, \( \chi ^{2} \) values given
in table 1 were not obtained using nominal errors provided by the photometric
data pipeline of Woźniak (2000). In some cases some error rescaling was needed
prior to the calculation of the fit. The reason for this is that image subtraction
method has minor problems with data error overestimation. Without rescaling
10 curves had their \( \chi ^{2} \) significantly smaller than 1. Error scaling
factor, determined for each curve separately, was defined as the ratio of flux
dispersion to average error. To compute the scaling factor we used only data
points from the baseline of the light curve. Baseline was defined as 75 percent
of all data points with smallest values of flux.

\begin{figure}[h]
{\par\centering \resizebox*{1\textwidth}{0.7\textheight}{\includegraphics{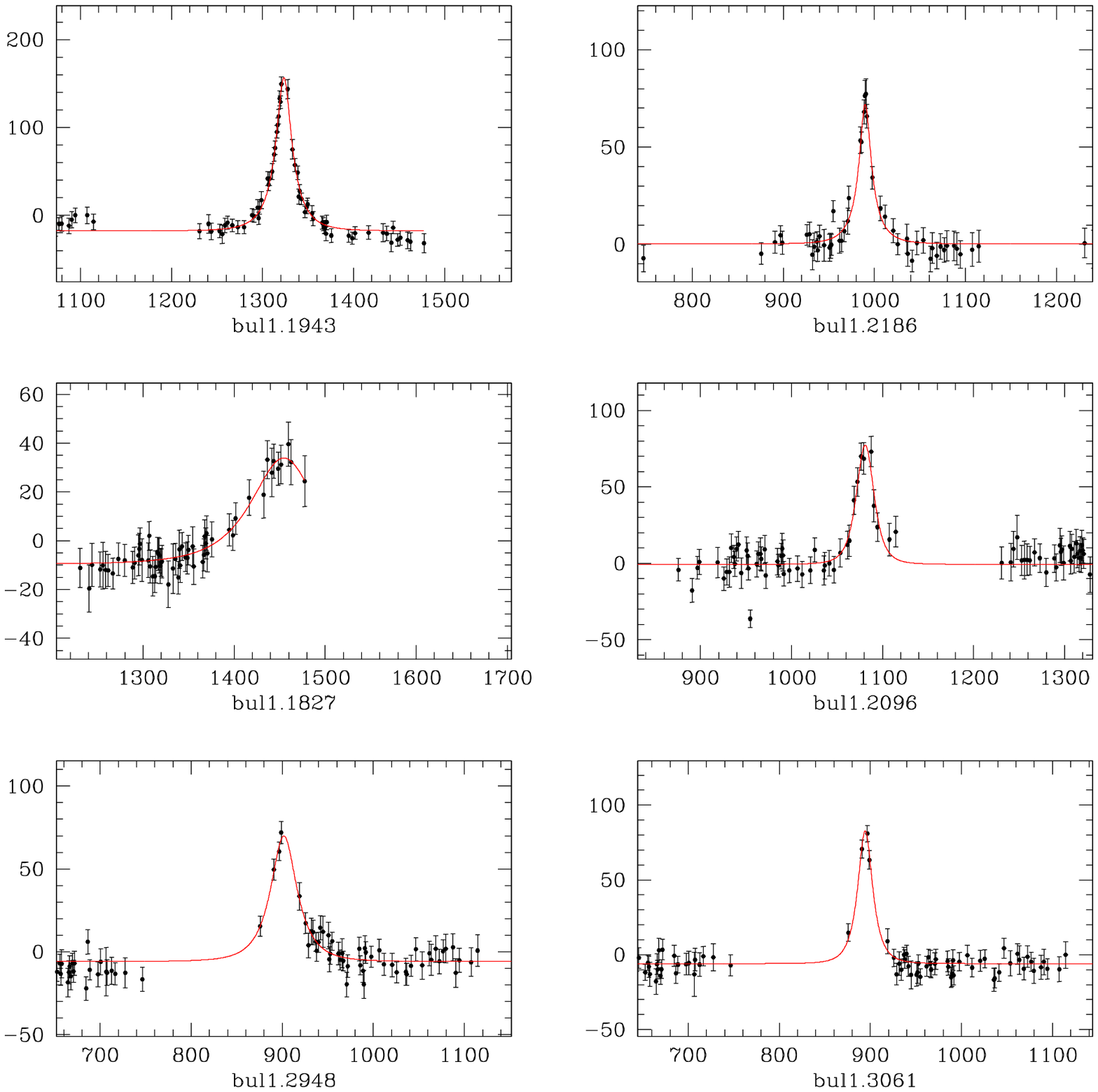}} \par}

\caption{{\small \label{lens1}Gravitational lensing events candidates found in OGLE
II bul\_sc1 image subtraction database}\small }
\end{figure}

\begin{figure}[h]
{\par\centering \resizebox*{1\textwidth}{0.7\textheight}{\includegraphics{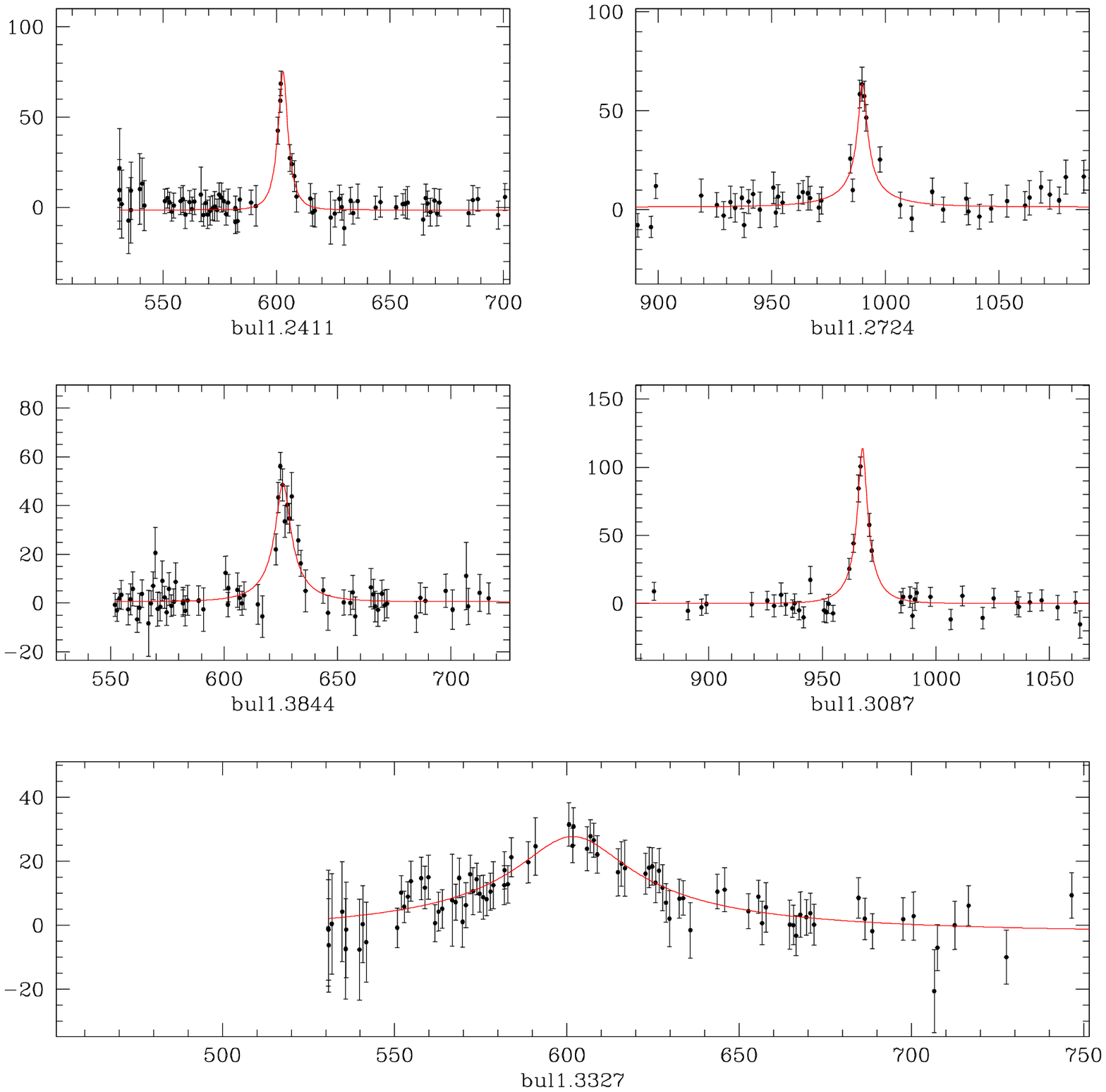}} \par}

\caption{{\small \label{lens2}Gravitational lensing events candidates found in OGLE
II bul\_sc1 image subtraction database (continued)}\small }
\end{figure}

Light curves of remaining 11 lenses are presented at Fig. \ref{lens1} and Fig.
\ref{lens2}. Though bul\_sc1.2948 and bul\_sc1.3061 may look similar, they
are separated on the reference frame by 347 pixels in x-coordinate and by 136
pixels in y-coordinate. This gives centroids distance about 372 pixels which
is quite large comparing to the size of the image and makes a hypothesis of
them being related to each other rather unlikely.

\section{Luminosity function and color-magnitude diagram}

We DoPhoted the reference frame in order to match star positions from our variable
star database with those found by DoPhot (for the detailed process of making
the reference image see Woźniak 2000). For every star in our database we sought
the closest DoPhot star within a 3 pixel radius. Due to the blending effect
some of the variable stars from the database were not detected by DoPhot. Almost
180 stars were not matched. Some others were identified with the same DoPhot
star. This problem concerns about 260 objects (there are 130 pairs of identically
matched stars). Most of them are stars with large proper motions, for details
see Eyer \& Woźniak (2001). 

\begin{figure}[h]
{\par\centering \resizebox*{10cm}{10cm}{\includegraphics{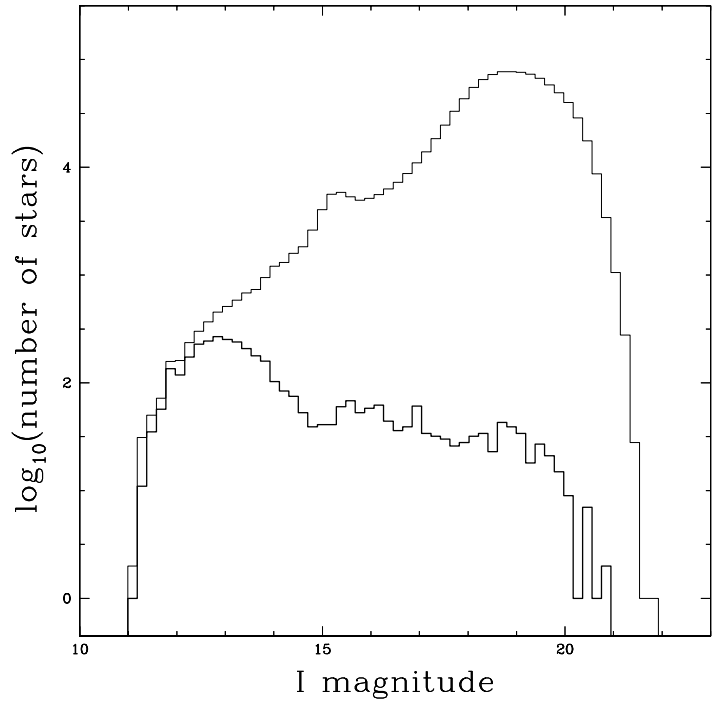}} \par}

{\par\centering \resizebox*{10cm}{10cm}{\includegraphics{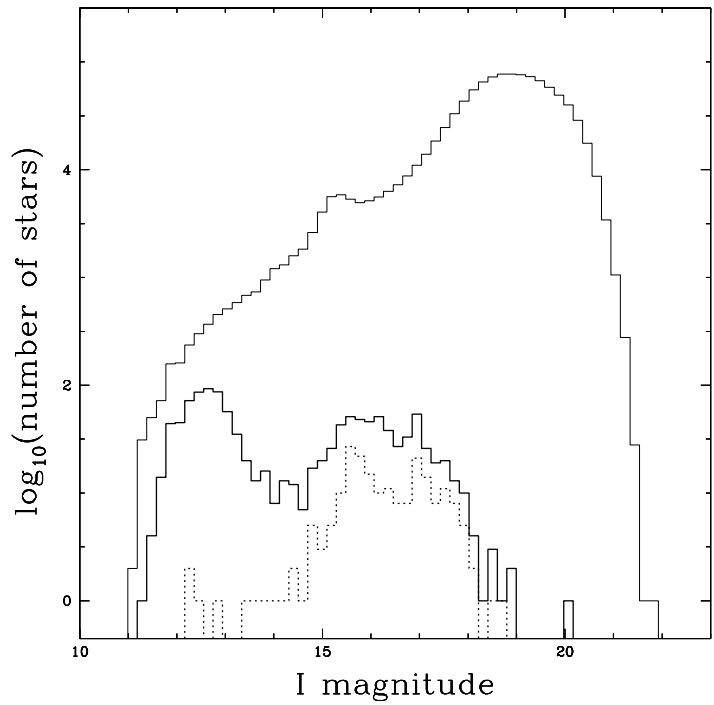}} \par}

\caption{{\small \label{lum}Thin line: Luminosity function (LF) in I magnitude. Upper
panel: thick line - LF for all variable stars. Lower panel: thick line - LF
for periodic stars with P < 50d, dotted line - W UMa, RR Lyrae and HADS variables.}\small }
\end{figure}

\begin{figure}[h]
{\par\centering \resizebox*{9cm}{9cm}{\includegraphics{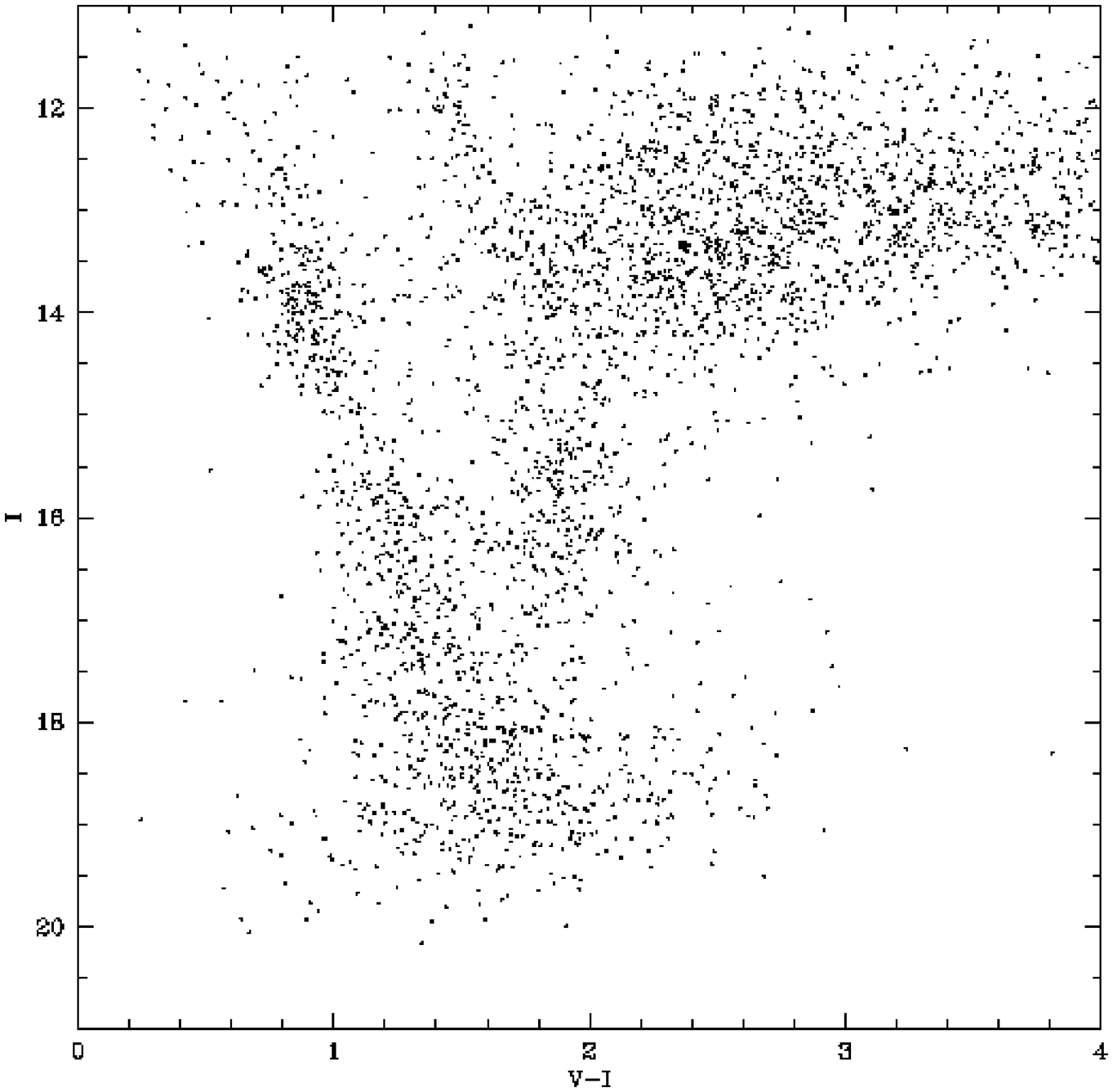}} \par}

{\par\centering \resizebox*{9cm}{9cm}{\includegraphics{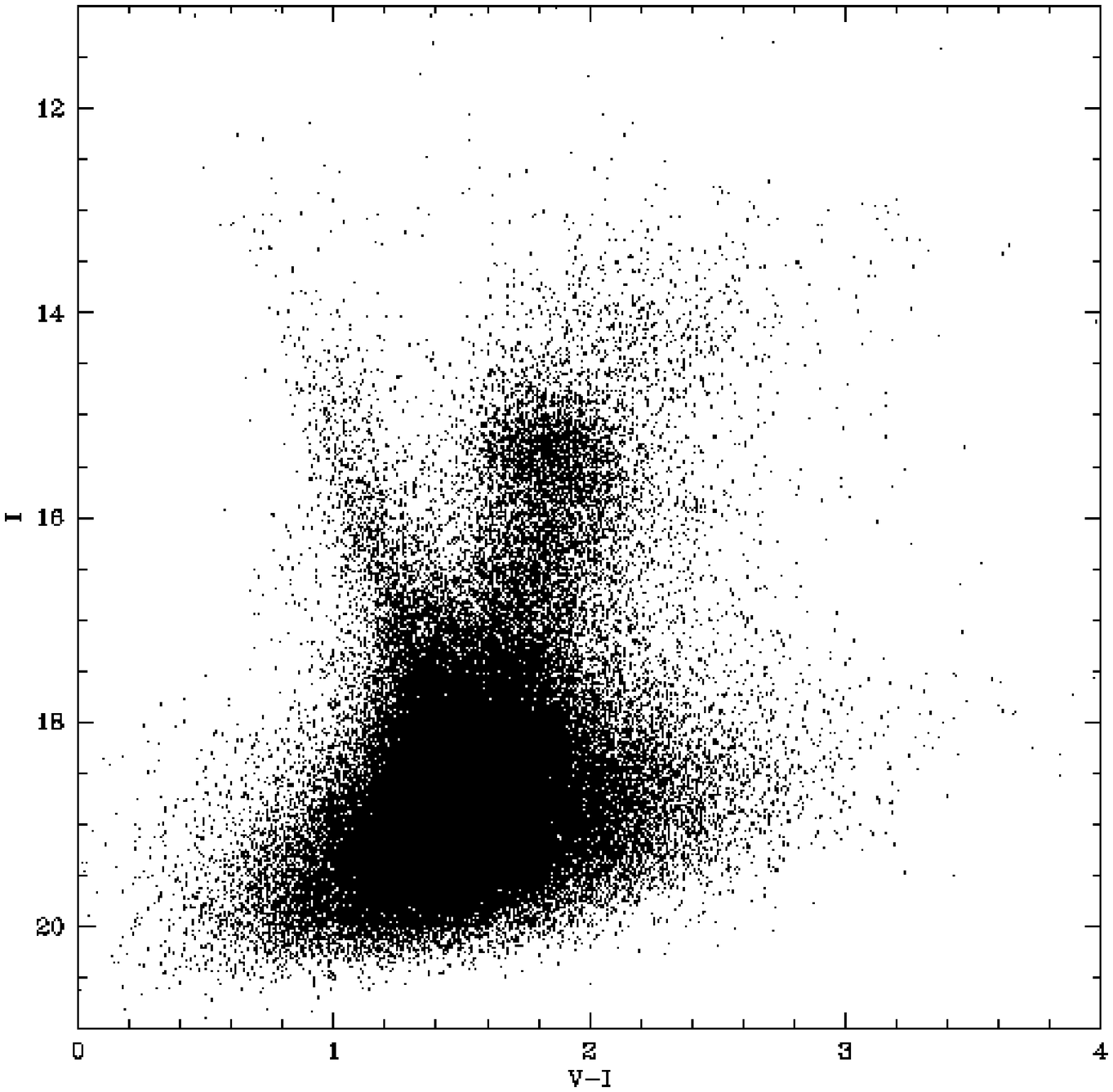}} \par}

\caption{{\small \label{lumcol}Color-magnitude diagram. Upper panel: variables. Lower
panel: whole field without variables (only one per 7 stars is shown).}\small }
\end{figure}

In Fig. \ref{lum} we show luminosity function for specific variability classes.
We also present the same function for the whole field, i.e. all stars found
by DoPhot. As one can see Fig. \ref{lum} suggests that almost all bright stars
are variable. Color-magnitude dependence for the field obtained by OGLE is shown
in Fig. \ref{lumcol} in the lower panel, in which all the variables were subtracted
and only one per every 7 stars was plotted for clarity. The upper panel contains
the same diagram for variable stars. Most of the brightest stars are red giants.
Group of Red Clump variables is dominated by stars with spots.

\begin{figure}[h]
{\par\centering \resizebox*{15cm}{15cm}{\includegraphics{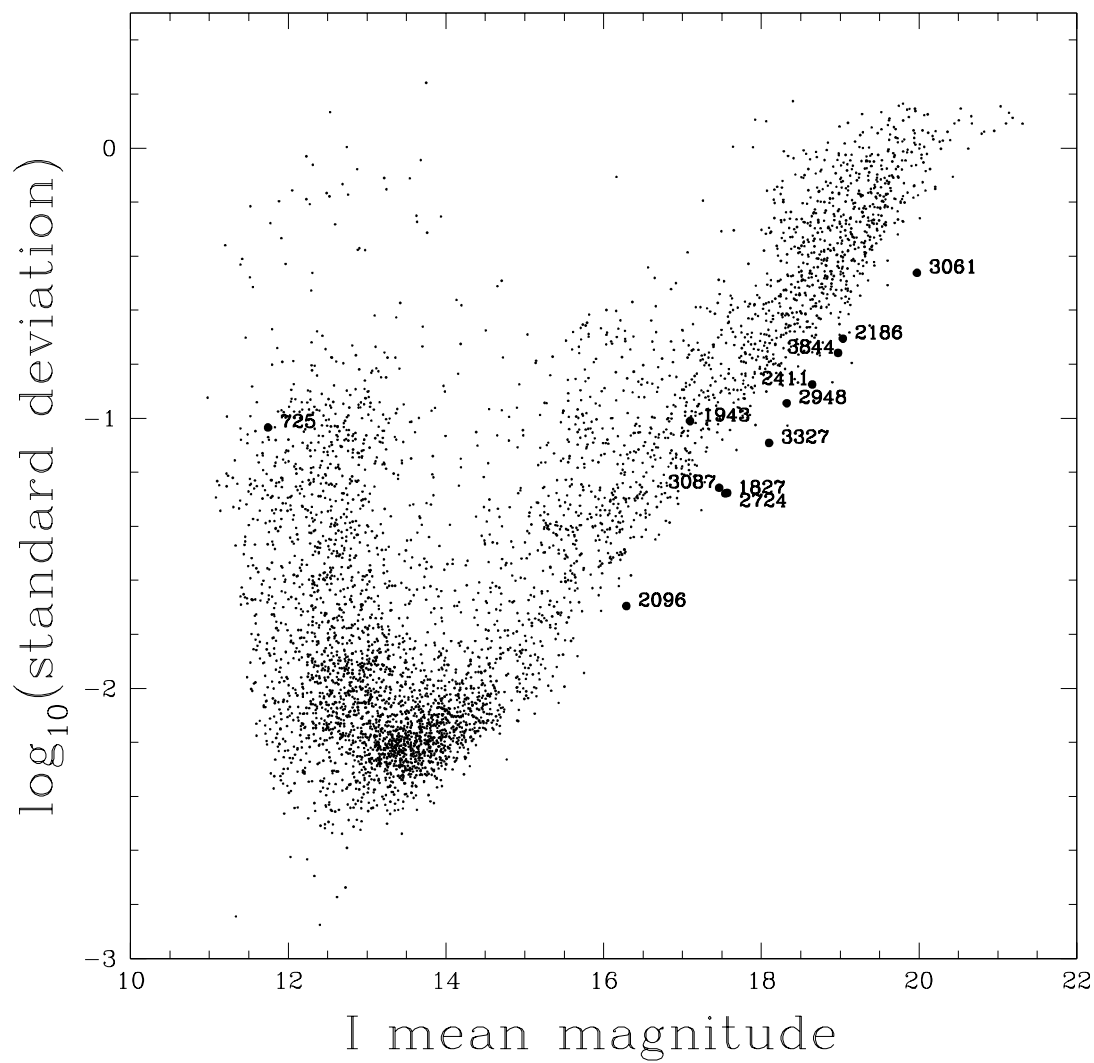}} \par}

\caption{{\small \label{magsig}Scale of variability described by \protect\( \log _{10}(\textrm{standard deviation})\protect \)
is shown against mean magnitude (in I filter) for all database variables. Filled
dots and corresponding numbers represents the gravitational lensing candidates
which are located along the line of detection limit.}\small }
\end{figure}

Our analysis confirms variability of most of the bright objects but in some
cases apparent variability may be due to the presence of saturated data points
or problems with the PSF. We rejected only obviously spurious variables. Also
notice that there are some non periodic variable stars with \( \mathrm{I}_{\mathrm{mag}}>20 \).
Some of them, but certainly not all, may in fact be spurious variables like
those described in section 2 but with correlation coefficients smaller than
0.75. Presence of some others may be due to different kinds of instrumental
defects. Nevertheless there are some real, weak, irregular variables with \( \mathrm{I}_{\mathrm{mag}}>20 \)
of rather unknown nature. The use of difference image analysis results in higher
number of specific variables found, e.g. gravitational lenses described in section
5.4. For example, during our search for episodic variables we found a strong
gravitational lens\emph{,} bul\_sc1.1943, which is located near another, brighter
star. This lens had not been detected previously with the use of DoPhot by Udalski
et al. (2000). It presumably merged with its bright neighbor`s PSF wings and
therefore was not identified. In Fig. \ref{magsig} we demonstrate variability
scale for variable stars of all kinds - note the position of gravitational lensing
candidates. They are located along the line of detection limit. Objects forming
very dense clump around 13 mag cover a wide range from blue to red stars in
Fig. \ref{lumcol}. The fluxes were converted to magnitudes as Woźniak (2000)
described.

\section{Summary and conclusions}

The database, derived by means of image subtraction method, contains only 4597
objects that were suspected of variability during the database generation process.
This value is 100 times smaller than the number of stars found on the reference
image by DoPhot. We classified 3969 of those 4597 stars, that is about 85\%,
as variables using the data pipeline described above. Our methods are complementary
and so a given star can pass more than one variability test. For example some
Mira type pulsators were not only found as periodic but also as episodic variables
as well. This was taken into account at the very end, after all the tests had
been performed. Final results of classification can be found at: 
\emph{ftp://ftp.astrouw.edu.pl/pub/mizerski/bul1.table}
in the form of the classification table presented below:

\( \,  \)

{\par\raggedright \texttt{\textbf{\footnotesize ~~~~ \#~~~~ ra:~~~~~~~~~~~~ dec:~~~~~~~~~
<Imag>: std\_dev:~~~ V-I:~~~ per:~ flag: }}\footnotesize \par}

\texttt{\textbf{\footnotesize ~~~~ 1~~~~ 18:02:01.07~~~~ -30:25:27.7~~~~ 16.034~~
0.001214~~~ -99.999~ 50.951 long}}{\footnotesize \par}

\texttt{\textbf{\footnotesize ~~~~ 2~~~~ 18:02:01.55~~~~ -30:25:07.2~~~~ 14.014~~
0.000046~~~ -99.999 -99.999 var}}{\footnotesize \par}

\texttt{\textbf{\footnotesize ~~~~ 3~~~~ 18:02:05.00~~~~ -30:25:25.1~~~~ 12.282~~
0.003312~~~~~ 2.206 -99.999 e}}{\footnotesize \par}

\texttt{\textbf{\footnotesize ~~~~ 4~~~~ 18:02:06.53~~~~ -30:25:02.7~~~~ 12.279~~
0.005765~~~~~ 4.338 -99.999 var}}{\footnotesize \par}

\texttt{\textbf{\footnotesize ~~~~ 5~~~~ 18:02:07.20~~~~ -30:25:07.3~~~~ 15.323~~
0.000792~~~~~ 1.927 -99.999 var}}{\footnotesize \par}

\texttt{\textbf{\footnotesize ~~~~ 6~~~~ 18:02:08.42~~~~ -30:25:42.9~~~~ 13.774~~
0.000077~~~~~ 3.086 -99.999 var}}{\footnotesize \par}

\texttt{\textbf{\footnotesize ~~~~ 7~~~~ 18:02:09.53~~~~ -30:25:15.1~~~~ 13.029~~
0.000617~~~~~ 4.436 -99.999 var}}{\footnotesize \par}

\texttt{\textbf{\footnotesize ~~~~ 8~~~~ 18:02:13.42~~~~ -30:25:52.0~~~~ 18.649~~
0.312950~~~~~ 1.842 -99.999 e}}{\footnotesize \par}

\texttt{\textbf{\footnotesize ~~~~ 9~~~~ 18:02:01.68~~~~ -30:25:35.6~~~~ 18.671~~
0.040266~~~ -99.999 -99.999 var}}{\footnotesize \par}

\texttt{\textbf{\footnotesize ~~~ 10~~~~ 18:02:03.95~~~~ -30:25:27.3~~~~ 14.253~~
0.000031~~~~~ 2.395 -99.999 var}}{\footnotesize \par}

\texttt{\textbf{\footnotesize ~~~ 11~~~~ 18:02:04.24~~~~ -30:25:52.8~~~~ 13.883~~
0.000041~~~~~ 2.289 -99.999 var}}{\footnotesize \par}

\texttt{\textbf{\footnotesize ~~~ 12~~~~ 18:02:07.07~~~~ -30:25:16.3~~~~ 17.639~~
0.021581~~~~~ 1.492~~ 0.249 wuma}}{\footnotesize \par}

\texttt{\textbf{\footnotesize ~}}{\footnotesize \par}

Details can be found in a \texttt{readme.txt} file on the web. 

Our methods let us detect 487 short periodic variables, with 71 RR Lyrae, 110
W UMa stars among them and 275 long periodic variables. We also detected 12
gravitational lenses candidates. The last result is very promising since with
the use of DoPhot only one lens, bul\_sc1.2186, had been previously found in
this field (Udalski et al. 2000). This is due to higher quality photometry and
different search methods: bul\_sc1.725\emph{,} which is possibly a lensing event
of a variable star, could not have been detected using constant stars catalogue.
On the other hand, the strongest lens bul\_sc1.1943 was not identified by DoPhot
as it presumably merged with another star PSF wings. Woźniak et al. (2001) detected
2 lenses in this field: bul\_sc1.1943 \emph{}and bul\_sc1.2186. Our others lenses,
with exception for bul\_sc1.3061 and \emph{}bul\_sc1.725, \emph{}are in their
``transient'' catalogue. 

We conclude that image subtraction method is much more efficient in massive
variability searches.

\paragraph{Acknowledgments}

First of all we would like to thank Dr. Bohdan Paczyński for all the advice
and helpful discussions and Dr. Wojciech Dziembowski, who carefully read the
script and gave us many valuable suggestions. We also thank Przemek Woźniak
who created the database and OGLE team at Warsaw for giving us access to their
data. 

This work was partly supported by the NSF grant AST-9820314 to Bohdan Paczyński.

\textbf{References }

{\small Alard, C., 1996, Astrophys. J., 458, L17}{\small \par}

{\small Alard, C., 2000, A\&AS, 144, 363}{\small \par}

{\small Alard, C., Lupton, R. H., 1998, Astrophys. J., 503, 325}{\small \par}

{\small Brandt, S., 1970, Statistical and Computational Methods in Data Analysis,
North-Holland Publishing Company}{\small \par}

{\small Eyer, L., Woźniak, P., 2001, astro-ph/0102027}{\small \par}

{\small Ruciński, S. M., 1993, PASP, 105, 1433}{\small \par}

{\small Schechter, P., Mateo, M., \& Saha, A., 1995, PASP, 105, 1342}{\small \par}

{\small Schwarzenberg - Czerny, A., 1989, Mon. Not. R. astr. Soc., 241, 153}{\small \par}

{\small Udalski, A., Kubiak, M., \& Szymański, M., 1997, Acta Astron., 47, 319}{\small \par}

{\small Udalski, A., Żebruń, K., Szymański, M., Kubiak, M., Pietrzyński, G.,
Soszyński, I., \& Woźniak, P., 2000, Acta Astron., 50, 1}{\small \par}

{\small Woźniak, P., 2000, Acta Astron., 50, 421 }{\small \par}

{\small Woźniak, P., Udalski, A., Szymański, M., Kubiak, M., Pietrzyński, G.,
Soszyński, I., Żebruń, K., 2001, arXiv:astro-ph/0106474v1}{\small \par}


\begin{figure}[h]
{\par\centering \resizebox*{0.9\textwidth}{0.65\textheight}{\includegraphics{rhopic.ps2}} \par}

\caption{{\small \label{per}Sample periodic light curves with different \protect\( \rho \protect \)
detected in the database.}\small }

{\par\centering {\small Difference flux is plotted against phase.}\small \par}
\end{figure}

\end{document}